\documentclass[runningheads]{llncs}
\usepackage{algorithmic,algorithm}
\usepackage{epsfig}
\begin{document}
\pagestyle{headings}
\titlerunning{{\textbf{A}utonomous \textbf{R}ange \textbf{T}ree}}
\authorrunning{S.~Sioutas et al.}

\title{ART : Sub-Logarithmic Decentralized Range Query Processing with Probabilistic Guarantees}

\author{S.~Sioutas\inst{1}
\and
P.~Triantafillou \inst{2}
\and
G.~Papaloukopoulos \inst{2}
\and \\
E.~Sakkopoulos \inst{2}
\and 
K.~Tsichlas \inst{3}
\and
Y.~Manolopoulos \inst{3}
}

\institute{Ionian University, Department of Informatics, sioutas@ionio.gr
\and
CTI and Dept.~of Computer Engineering \& Informatics, University of Patras, (peter, papaloukg, sakkopul)@ceid.upatras.gr
\and
Aristotle University of Thessaloniki, Department of Informatics,
(tsichlas, manolopo)@csd.auth.gr}

\maketitle

\begin{abstract}
We focus on range query processing on large-scale, typically distributed infrastructures, such as clouds of thousands of nodes of shared-datacenters, of p2p distributed overlays, etc. In such distributed environments, efficient range query processing is the key for managing the distributed data sets per se, and for monitoring the infrastructure's resources. We wish to develop an architecture that can support range queries in such large-scale decentralized environments and can scale in terms of the number of nodes as well as in terms of the data items stored. Of course, in the last few years there have been a number of solutions (mostly from researchers in the p2p domain) for designing such large-scale systems. However, these are inadequate for our purposes, since at the envisaged scales the classic logarithmic complexity (for point queries) is still too expensive while for range queries it is even more disappointing. In this paper 
\footnote{A limited and preliminary version of this work has been presented as brief announcement in Twenty-Ninth Annual ACM SIGACT-SIGOPS Symposium on Principles of Distributed Computing, Zurich, Switzerland July 25-28, 2010 \cite{SPSTMT10}} we go one step further and achieve a sub-logarithmic complexity. We contribute the ART \footnote{\textbf{A}utonomous \textbf{R}ange \textbf{T}ree}
 structure, which outperforms the most popular decentralized structures, including Chord (and some of its successors), BATON (and its successor) and Skip-Graphs. We contribute theoretical analysis, backed up by detailed experimental results, showing that the communication cost of query and update operations is $O(\log_{b}^2 \log N)$ hops, where the base $b$ is a double-exponentially power of two and $N$ is the total number of nodes. Moreover, ART is a fully dynamic and fault-tolerant structure, which supports the join/leave node operations in $O(\log \log N)$ expected w.h.p number of hops. 
Our experimental performance studies include a detailed performance comparison which showcases the improved performance, scalability, and robustness of ART.
\end{abstract}
\centerline{{\bf Keywords}:Distributed Data Structures, P2P Data Management.}

\section{Introduction and Motivation}

Decentralized range query processing is a notoriously difficult problem to solve efficiently and scalably in decentralized network infrastructures. It has been studied in the last years extensively, particularly in the realm of p2p, which is increasingly used for content delivery among users. There are many more real-life applications in which the problem also materializes. Consider the (popular nowadays) cloud infrastructures for content delivery.
Monitoring of thousand of nodes, where thousands of different applications from different organizations execute, is an apparent requirement. This monitoring process often requires support for range queries over this decentralized infrastructure:  consider range queries that are issued in order to identify which of the cloud nodes are under-utilized, (i.e., $utilization < threshold$) in order to assign to them more data \& tasks and better exploit all available resources, increasing the
revenues of the cloud infrastructure, or to identify overloaded nodes, ($load > threshold$) in order to avoid bottlenecks in the cloud, which hurts overall performance, and revenues.
 
Each node in the cloud maintains a tuple with attributes: utilization, OS, load, NodeId, e.t.c. Collectively,
these makeup a relation, CloudNodes, and we wish to execute queries such as:\\
SELECT NodeId\\
FROM Cloudnodes\\
WHERE low $<$ utilization $<$ high

or point and range queries, e.g.\\
SELECT NodeId\\
FROM Cloudnodes\\
WHERE low $<$ utilization $<$ high and OS=UNIX

An acceptable solution for processing range queries in such large-scale decentralized environments must scale in terms of the number of nodes as well as in terms of the number of data items stored. The available solutions for architecting such large-scale systems are inadequate for our purposes, since at the envisaged scales (trilions of data items at millions of nodes) the classic logarithmic complexity (for point queries) offered by these solutions is still too expensive. And for range queries, it is even more disappointing. Further, all available solutions incur large overheads with respect to other critical operations, such as join/leave of nodes, and insertion/deletion of items.
Our aim with this work is to provide a solution that is comprehensive and outperforms related work {\it with respect to all major operations, such as lookup, join/leave, insert/delete, and to the required routing state that must be maintained in order to support these operations}. Specifically, we aim at achieving a sub-logarithmic complexity for all the above operations!

Peer-to-peer (P2P) systems have become very popular, in both academia and industry. They are widely used for sharing resources like music files etc. Search for a given ID, is a crucial operation in P2P systems, and there has been considerable recent work in devising effective distributed search (a.k.a. lookup) techniques. The proposed structures include a ring as in Chord \cite{KKSMB01}, a multiple dimensional grid as in CAN \cite{RFHKS01}, a multiple list as in SkipGraph \cite{AS03,GNS06}, or a tree as in PHT \cite{RRHS04}, BATON \cite{JOV05} and BATON* \cite{JOTVZ06}. Most search structures (including all the ones just mentioned except for BATON* and PHT) bound the search cost to a base 2 logarithm of the search space: for a system with $N$ peer nodes, the search cost is bounded by $O(\log N)$. Relative to tree-based indexes, a disadvantage of PHTs (Prefix Hash Trees) is that their complexity is expressed in terms of the $\log$ of the domain size, $D$, rather than the size of the data set, $N$ and depends on distribution over $D-bit$ keys. BATON* is a multi-way search tree, which reduces the search cost to $O(\log_m N)$, where $m$ is the tree fanout. The penalty paid is a larger update cost, but no worse than linear in $m$.
One of the distributed indexes with high fanout is the P-Tree \cite{CLGS04}, where each peer maintains a B$^+$-tree leaf and a path of virtual index nodes from the root to the specific leaf. Search is very effective, but updates are expensive, possibly requiring substantial synchronization effort.
BATON* extends BATON by allowing a fanout of $m>2$. Thus, the search cost becomes $O(\log_m N)$, as expected. Moreover, the cost of updating routing tables is $O(m·\log_m N)$ only, as compared to $O(\log_2 N)$ in BATON - an improvement that is better than linear in $m$.
Furthermore, BATON* has better fault tolerance properties than BATON, and supports load balancing more efficiently. In fact, the system's fault tolerance, measured as the number of nodes that must fail before the network is partitioned, increases linearly with $m$. Similarly, the expected cost of load balancing decreases linearly with $m$. 

\textbf{Our Results:} In this paper we present the ART structure, which outperforms the most popular decentralized structures, including Chord (and some of its successors), BATON and BATON* and Skip-Graphs. ART is an exponential-tree structure, which remains unchanged w.h.p., and organizes a number of fully-dynamic buckets of peers. We provide and analyze all relevant algorithms for accessing ART. We contribute theoretical analysis, backed up by detailed experimental results, showing that the communication cost of query and update operations is $O(\log_{b}^2 \log N)$ hops, where the base $b$ is a double-exponentially power of two. Moreover, ART is a fully dynamic and fault-tolerant structure, which supports the join/leave node operations in $O(\log \log N)$ expected w.h.p number of hops. Since ART is a tree based system, our experimental performance studies include our development of BATON* (the best current tree based system), 
and a detailed performance comparison which showcases the improved performance, scalability, and robustness of ART.

In Section 2 we present more thoroughly key previous work. Section 3 describes the ART structure and analyzes its basic functionalities. Section 4 presents a thorough experimental evaluation; Section 5 presents some interesting heuristics and thresholds, whereas Section 6 concludes the paper. 

\section{Previous Work}

Existing structured P2P systems can be classified into three categories: distributed hash table (DHT) based systems, skip list based systems, and tree based systems.
There are several P2P DHT architectures like Chord \cite{KKSMB01}, CAN \cite{RFHKS01}, Pastry \cite{RD01}, Tapestry \cite{ZHSRJK04}, Kademlia \cite{MM02} and and Kelips \cite{GBLDR03}. Unfortunately, these systems cannot easily support range queries since DHTs destroy data ordering. 
This means that they cannot support common queries such as "find all research papers published from 2004 to 2008". To support range queries, inefficient DHT variants have been proposed (for details see \cite{GAA03}, \cite{SGAA04}, \cite{AX02}, \cite{TP03}). 


Skip list based systems such as Skip Graph \cite{AS03,GNS06} and Skip Net \cite{HJSTW03} are based on the skip-list structure. To provide decentralization they use randomized techniques to create and maintain the structure. Moreover, they can support both exact match queries and range queries by partitioning data into ranges of values. However, they cannot guarantee data locality (which hurts efficient range query processing) and load balancing in the system. 

Tree based systems also carry their own disadvantages. P-Grid \cite{CLGS04} utilizes a binary prefix tree. It can neither guarantee the bound of search steps since it cannot control the tree height. An arbitrary multi-way tree was proposed in \cite{LNSTB04}, where each node maintains links to its parent, children, sibling and neighbors. It also suffers from the same problem. P-Tree \cite{CLGS04} utilizes a B$^+$-tree on top of the CHORD overlay network, and peers are organized as a CHORD ring, each peer maintaining a data leaf and a left most path from the root to that B$^+$-tree node. This results in significant overhead in building and maintaining the consistency of the B$^+$-tree. In particular, a tree has been built for each joining node, and periodically, peers have to exchange their stored B$^+$-tree for checking consistency. BATON \cite{JOV05} utilizes a binary balanced tree and as a consequence, it can control the tree height and avoid the problem of P-Grid. Nevertheless, similarly to other P2P systems, BATON's search cost is bounded by $O(\log_2 N)$. BATON* \cite{JOTVZ06} is an overlay multi-way tree based on B-trees, with better searching performance. The penalty paid is a marginally larger update cost. 

Systems like MAAN \cite{CFCS03}, Mercury \cite{BAS04} and DIM \cite{LKGH03} support multi-attribute queries in a multi-dimensional space. BATON* can also effectively support queries over multiple attributes. In addition to supporting the use of multiple attributes in a single index, BATON* further introduces the notion of attribute classification, based on the importance of the attribute for querying, and the notion of attribute groups. In particular, BATON* relies on the construction of multiple independent indexes for groups of one or more attributes. For further details about the suggested techniques for partitioning attributes into such groups, see \cite{JOTVZ06}.

\begin{table*}[!bht]
\begin{center}
\begin{tabular}{|c|c|c|c|} \hline\hline 
\small{P2P} & \small{Lookup, Insert, }  & \small{Maximum Size} & \small{Join/} \\ 
\small{architectures} & \small{Delete key}  & \small{of routing table} & \small{Depart peer} \\ \hline\hline 
\small{CHORD} & $\small{O(\log N)}$ & $\small{O(\log N)}$ & $\small{O(\log N)}$ \small{w.h.p.} \\ \hline
\small{H-F-Chord(a)} & $\small{O(\log N/ \log \log N)}$ & $\small{O(\log N)}$ & $\small{O(\log N)}$ \\ \hline
\small{LPRS-Chord} & $\small{O(\log N)}$ & $\small{O(\log N)}$ & $\small{O(\log N)}$ \\ \hline
\small{Skip Graphs} & $\small{O(\log N)}$ & $\small{O(1)}$ & $\small{O(\log N)}$ \small{amortized} \\ \hline
\small{BATON} & $\small{O(\log N)}$ & $\small{O(\log N)}$ & $\small{O(\log N)}$ \small{w.h.p.} \\ \hline
\small{BATON*} & $\small{O(\log_m N)}$ & $\small{O(m\log_m N)}$ & $\small{O(m \log_m N)}$ \\ \hline
\small{ART-tree} & $\small{O(\log_{b}^2 \log N)}$ & $\small{O(N^{1/4}/ \log^c N)}$ & $\small{O(\log \log N)}$ \small{expected w.h.p.} \\ \hline\hline 
\end{tabular}
\end{center}
\caption{\small{Performance comparison between ART, Chord, BATON and Skip Graphs.}}
\label{tab:perf_comp}
\end{table*}

For comparison purposes, in Table 1 we present a qualitative evaluation with respect to elementary operations between ART, Skip-Graphs, Chord and its newest variations (F-Chord($\alpha$) \cite{SMLKKDB03}, LPRS-Chord \cite{ZGG03}), BATON \cite{JOV05} and its newest variation BATON* \cite{JOTVZ06}. It is noted that $c$ is a big positive constant. 

\section{Our Solution}

First, we build the LRT (\textbf{L}evel \textbf{R}ange \textbf{T}ree) structure, one of the basic components of the final ART structure. LRT will be called upon to organize collections of peers at each level of ART.\\

\subsection{Building LRT structure}
\begin{figure}[htbp]
	\centering
		\includegraphics[width=0.70\textwidth]{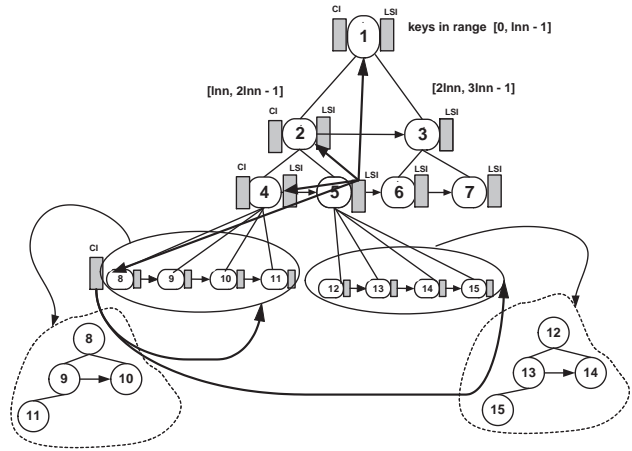}
	\caption{The LRT structure for b=2}
	\label{fig:LRT}
\end{figure}

LRT is built by grouping peers having the same ancestor and organizing them in a tree structure recursively. The innermost level of nesting (recursion) will be characterized by having a tree in which no more than $b$ peers share the same direct ancestor, where $b$ is a double-exponentially power of two (e.g. 2,4,16,...). Thus, multiple independent trees are imposed on the collection of peers. Figure~\ref{fig:LRT} illustrates a simple example, where $b=2$. 

The degree of the peers at level $i>0$ is $d(i)=t(i)$, where $t(i)$ indicates the number of peers at level $i$. It holds that $d$(0)=b and $t$(0)=1. 
Let $n$ be $w$-bit keys. Each peer with label $i$ (where $1 \leq i \leq N$) stores ordered keys that belong in the range [$(i-1) \ln n , i \ln n$--1], where $N=n/lnn$ is the number of peers. Note here that the $lnn$ (and not $logn$) factor is due to a specific combinatorial game (\cite{KMSTTZ03}) we invoke in the next subsection. 

We also equip each peer with a table named {\it Left Spine Index} (LSI), which stores pointers to the peers of the left-most spine (see pointers starting from peer 5). 

Furthermore, each peer of the left-most spine is equipped with a table named {\it Collection Index} (CI), which stores pointers to the collections of peers presented at the same level (see pointers directed to collections of last level). Peers having the same father belong to the same collection. For example, in Figure~\ref{fig:LRT}, peers 8, 9, 10, and 11 constitute a certain collection.

\subsubsection{Lookup Algorithm}
Assume we are located at peer $s$ (we mean the peer labeled by integer number $s$) and seek a key $k$. First, we find the range where $k$ belongs in. Let say $k \in [(j-1)$ $\ln n, j \ln n-1]$. The latter means that we have to search for peer $j$. The first step of our algorithm is to find the LRT level where the desired peer $j$ is located. For this purpose, we exploit a nice arithmetic property of LRT. This property says that for each peer $x$ located at the left-most spine of level $i$, the following formula holds:
\begin{equation} \label{eq:father}
label(x)=label(father(x))+b^{2^{i-2}}
\end{equation}
For example, peer 4 is located at level 2, thus $4=father(4)+2$ or peer 8 is located at level 3, thus $8=father(8)+4$ or peer 24 (not depicted in the Figure~\ref{fig:LRT}) is located at level 4, thus $24=father(24)+16$. The last equation is true since $father(24)=8$.

Thus, for each level $i$ (in the next subsection we will prove that $0 \leq i \leq \log \log N$), we compute the label $x$ of its left most peer by applying Equation~(\ref{eq:father}). Then, we compare the label $j$ with the computed label $x$. If $j \geq x$, we continue by applying Equation (1), otherwise we stop the loop process with current value $i$. The latter means that peer $j$ is located at the $i$-th level. So, first we follow the $i$-th pointer of the LSI table located at peer $s$ so as to reach the leftmost peer $x$ of level $i$. Then, we compute the collection in which the peer $j$ belongs. Since the number of collections at level $i$ equals the number of peers located at level $(i-1)$, we divide the distance between $j$ and $x$ by the factor $t(i-1)$. Let $m$ (in particular $m=\left\lceil \frac{j-x+1}{t(i-1)} \right\rceil$) be the result of this division. The latter means that we have to follow the $(m+1)$-th pointer of the CI table so as to reach the desired collection. Since the collection indicated by the CI[$m$+1] pointer is organized in the same way at the next nesting level, we continue this process recursively.

\subsubsection{Analysis}
The degree of the peers at level $i>0$ is $d(i)=t(i)$, where $t(i)$ indicates the number of peers at level $i$. It is defined that $d$(0)=b and $t$(0)=1. It is apparent that $t(i)=t(i-1)d(i-1)$, and, thus, by putting together the various components, we can solve the recurrence and obtain $d(i)=t(i)=b^{2^{i-1}}$ for $i\geq 1$. This double exponentially increasing fanout guarantees the following lemma:

\noindent {\bf Lemma 1}: The height (or the number of levels) of LRT is $O(\log \log_{b} N)$ in the worst case.

The size of the LSI table equals the number of levels of LRT. Moreover, the maximum size of the $CI$ table appears at last level. It is apparent from the building of the LRT structure that at last level $h$, $t(h)=O(N)$. It holds that $t(h)=b^{2^{h-1}}$, thus $b^{2^{h-1}}=O(N)$ or $h-1=O(loglog_{b}N)$ or $h=O(loglog_{b}N)+1$. Since the number of collections at level $h$ equals the number of peers located at level $(h-1)$ we take $t(h-1)=b^{2^{h-2}}=b^{2^{(O(loglog_{b}N)+1)-2}}$ or
$b^{2^{O(loglog_{b}N)-1}}=b^{2^{O(loglog_{b}N)}2^{-1}}=\left(b^{2^{O(loglog_{b}N)}}\right)^{1/2}$ and the lemma 2 follows:
 
\noindent {\bf Lemma 2}: The maximum size of the $CI$ and $LSI$ tables is $O(\sqrt{N})$ and
$O(\log \log N)$ in worst-case respectively.

We need now to determine what will be the maximum number of nesting trees that can occur for $N$ peers. Observe that the maximum number of peers with the same direct ancestor is $d(h-1)$. Would it be possible for a second level tree to have the same (or bigger) depth than the outermost one? 

This would imply that $\sum_{j=0}^{h-1}t(j)<d(h-1)$.

As otherwise we would be able to fit all the $d(h-1)$ peers within the first $h-1$ levels. But we need to remember that $d(i)=t(i)$, thus $d(h-1)+\sum_{j=0}^{h-2}d(j)<d(h-1)$.

This would imply that the number of peers in the first $h-2$ levels is negative, clearly impossible. Thus, the second level tree will have depth strictly lower than the depth of the outermost tree.

The innermost (let say $j^{th}$) level of nesting (recursion) is characterized by having a tree in which no more than $b$ nodes share the same direct ancestor, where $b$ is a double-exponentially power of two (e.g. 2,4,16,...). In this case $b=N^{1/b^{j}}$ and the lemma 3 follows:

\noindent {\bf Lemma 3}: The maximum number of possible nestings in LRT structure is $O(\log_{b} \log N)$ in the worst case.

At each peer we pay an extra processing cost by repeating the equation~(\ref{eq:father}) $O(\log \log N)$ times at most in order to locate the desired LSI pointer. Then, we need $O(1)$ hops for locating the left-most peer $x$ of the desirable level. We must note here that the processing overhead compared to communication overhead is negligible, thus we can ignore the $O(\log \log N)$ processing factor at each peer. Finally we need $O(1)$ hops for locating the desirable collection of peers via the CI[$m$+1] pointer. Since, the collection indicated by the CI[$m$+1] pointer is organized in the same way at a next nesting level, we continue the above process recursively. According to lemma 2 the maximum number of nesting levels is $O(\log_{b} \log N)$, and the theorem follows:

\noindent {\bf Theorem 1}: Exact-match queries in the LRT structure require $O(\log_{b} \log N)$ hops or lookup messages in the worst case.

\subsection{Building ART Structure}

\begin{figure*}[bht]
\begin{center}
\includegraphics[width=0.90\textwidth]{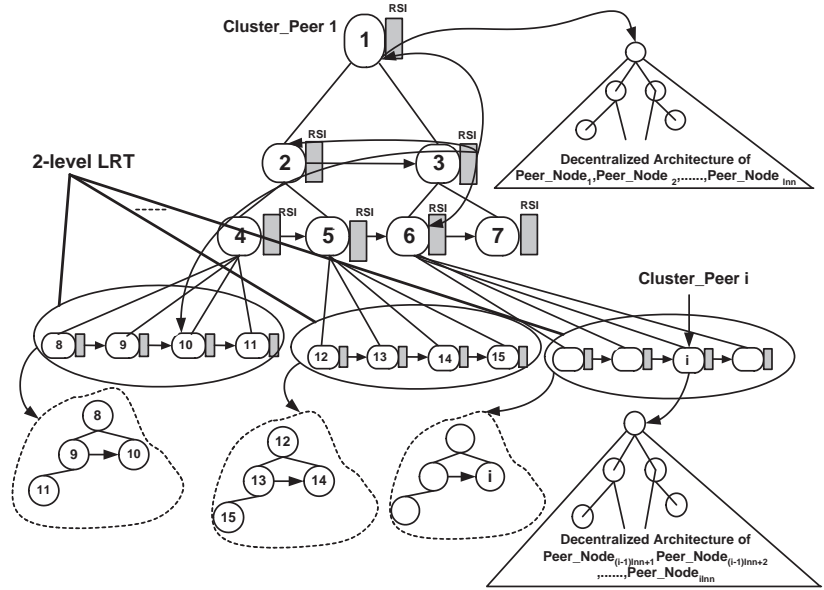}
\end{center}
\caption{The ART structure for b=2}
\label{fig:ART}
\end{figure*}

\begin{figure*}[bht]
\begin{center}
\includegraphics[width=0.90\textwidth]{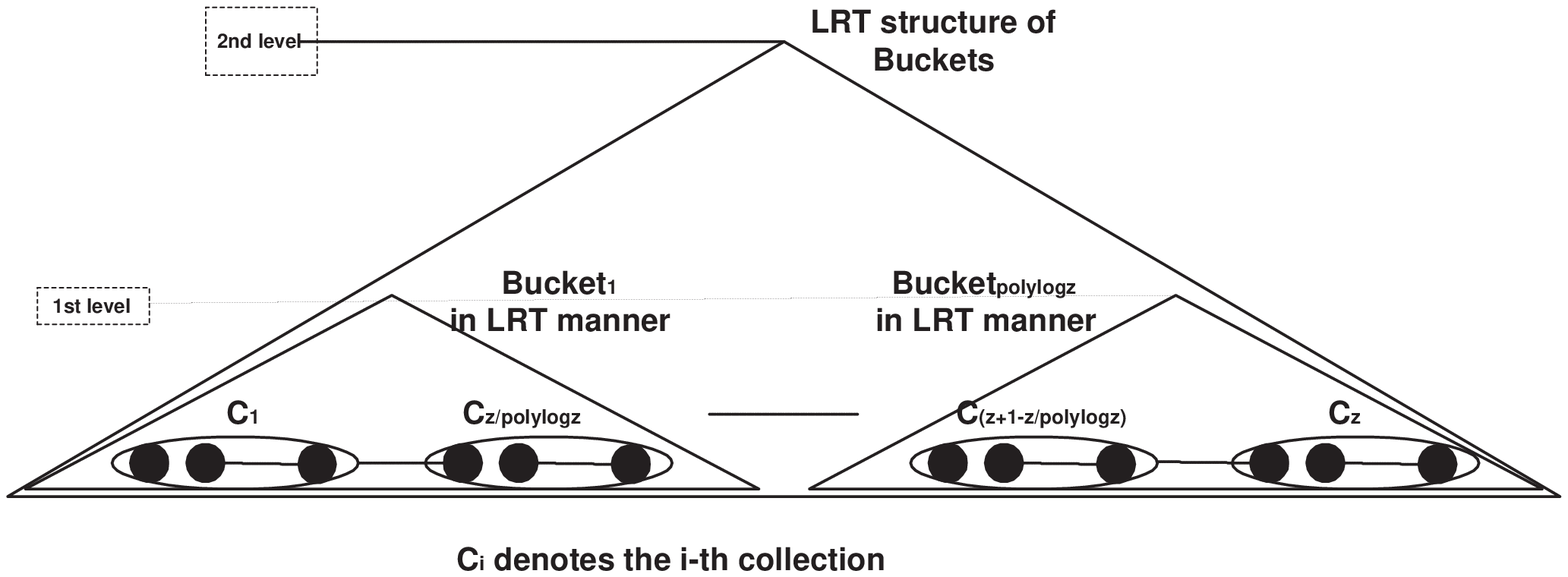}
\end{center}
\caption{The 2-level LRT structure}
\label{fig:2levelLRT}
\end{figure*}

We define as cluster\_peer a bucket of $\Theta({\rm polylog}~N')$ ordered peers, where $N'$ is the number of cluster\_peers.

At initialization step we choose as cluster\_peers the 1st peer, the $(\ln n +1)$-th peer, the $(2 \ln n +1)$-th peer and so on. This means that each cluster\_peer with label $i'$ (where $1 \leq i'\leq N'$) stores ordered peers with sorted keys belonging in the range $[(i'-1) \ln^2 n, \ldots, i' \ln^2 n-1]$, where $N'=n/ \ln^2 n$ is the number of cluster\_peers. 

ART stores cluster\_peers only, each of which is structured as an independent decentralized architecture. The backbone-structure of ART is exactly the same with LRT (see Figure~\ref{fig:ART}). Moreover, instead of the {\bf L}eft-most {\bf S}pine {\bf I}ndex (LSI), which reduces the robustness of the whole system, we introduce the {\bf R}andom {\bf S}pine {\bf I}ndex (RSI) routing table, which stores pointers to randomly chosen (and not to left-most) cluster\_peers (see in Figure~\ref{fig:ART} the pointers starting from peer 3). In addition, instead of using fat $CI$ tables, we access the appropriate collection of cluster\_peers by using a 2-level LRT structure. The 2-level LRT is an LRT structure over $\log^{2c}Z$ buckets each of which organizes $\frac{Z}{\log^{2c}Z}$ collections in a LRT manner, where $Z$ is the number of collections at current level and $c$ is a big positive constant (see Figure~\ref{fig:2levelLRT}) 

\subsubsection{Load Balancing}

We model the join/leave of peers inside a cluster\_peer as the combinatorial game of bins and balls presented in \cite{KMSTTZ03} and the lemma 4 follows:

{\bf Lemma 4}: Given a $\mu(\cdot)$ random sequence of join/leave peer operations, the load of each cluster\_peer never becomes zero and never exceeds $\Theta({\rm polylog}~N')$ size in expected w.h.p. case.

\subsubsection{Routing Overhead}

ART stores cluster\_peers, each of which is structured as an independent decentralized architecture (be it BATON*, Chord, Skip-Graph, e.t.c.) (see Figure~\ref{fig:ART}). Here, we will try to avoid the existence of CI routing tables, since these tables may become very large ($O(\sqrt{N})$) in the worst case as well as the occurrence of local hot spots in the left-most spine results in a less robust decentralized infrastructure. Thus, instead of the {\bf L}eft-most {\bf S}pine {\bf I}ndex (LSI), we introduce the {\bf R}andom {\bf S}pine {\bf I}ndex (RSI) routing table. The latter table stores pointers to the cluster\_peers of a random spine (for example, in Figure~\ref{fig:ART} the randomly chosen cluster\_peers 1, 2, 6 and 10 are pointed to by the RSI table of cluster\_peer 3). Furthermore, instead of CI tables, we can access the appropriate collection of cluster\_peers by using the 2-level LRT structure discussed above (see Figure~\ref{fig:2levelLRT}). Since the larger number of collections is $Z=O(N^{1/2})$ (it appears in the last level), the overhead of routing information is dominated by the second level structures in each of which we have an $O(\sqrt{\frac{Z}{\log^{2c}Z}})=O(N^{1/4}/ \log^c N)$ routing overhead. Thus, Theorem 2 follows: 

\noindent {\bf Theorem 2}: The overhead of routing information in ART is $O(N^{1/4}/ \log^c N)$ in the worst case.\\
\noindent {\bf Remark 1}: If we use a k-level LRT structure, the routing information overhead becomes $O(N^{1/2^{k}}/ \log^c N)$ in the worst case.

\subsubsection{Lookup Algorithms}

\begin{figure*}[bht]
\begin{center}
\includegraphics[width=0.80\textwidth]{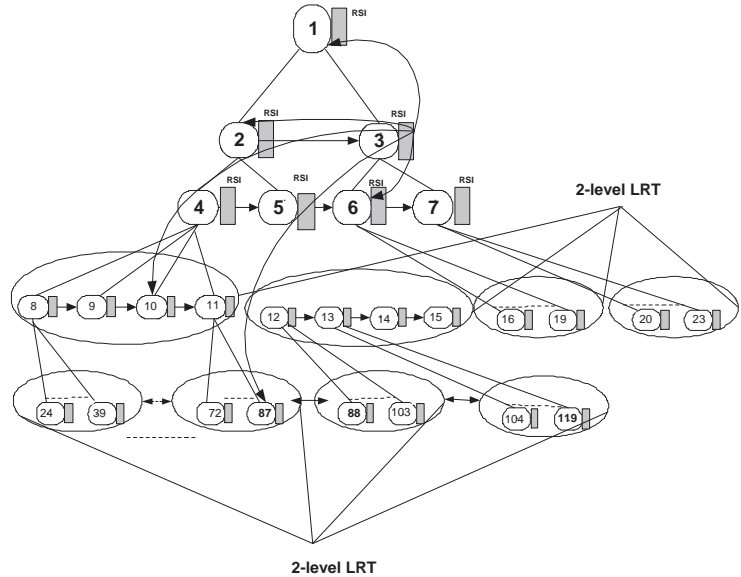}
\end{center}
\caption{An example of Lookup Steps via RSI[ ] tables and 2-level LRT structures}
\label{fig:case3}
\end{figure*}

Let us explain the lookup operations in ART. For example, in Figure~\ref{fig:case3} suppose we are located at cluster\_peer 3 and we are looking for two keys, which are located at cluster\_peers 19 and 119 respectively. The first step of our algorithm is to find the levels of the ART where the desired cluster\_peers (e.g. 19 and 119) are located. In our example, the fourth and fifth levels are the desired levels. By following the RSI[4] and RSI[5] pointers we reach the cluster\_peers 10 and 87 respectively. Now, we are starting from peers 10 and 87 to lookup the peers 19 and 119 respectively in the 2-level LRT structures of the collections in respective levels.\\
Generally speaking, since the maximum number of nesting levels is $O(\log_{b} \log N)$ and at each nesting level $i$ we have to apply the standard LRT structure in $N^{1/2^i}$ collections, the whole searching process requires $T_1(N)$ hops or lookup messages to locate the target cluster\_peer, where:
\begin{equation}
T_1(N) = \sum_{i=0}^{\log_{b} \log N} \log_{b} \log (N^{1/2^i}) = 
\log_{b} (\prod_{i=0}^{\log_{b} \log N} \log (N^{1/2^i}))
\end{equation}
where
$$\prod_{i=0}^{\log_{b} \log N} \log (N^{1/2^i}) < (\log{N})^{\log_b\log{N}}$$
from which we get:
\[
T_1(N) < \log_{b}((\log{N})^{\log_{b} \log N}) = O(\log_{b}^2 \log N)
\]
Then, we have to locate the target peer by searching the respective decentralized structure, requiring $T_2(N)$ hops. Since each of the known decentralized architectures requires a logarithmic number of hops, the total process requires $T(N)=T_1(N)+T_2(N)=O(\log_{b}^2 \log N)$ hops or lookup messages and the theorem follows.

\noindent {\bf Theorem 3}: Exact-match queries in the ART structure require $O(\log_{b}^2 \log N)$ hops or lookup messages. 

Having located the target peer for key $k_\ell$ and exploiting the order of keys on each node, range queries of the form $[k_\ell, k_r]$ require an $O(\log_{b}^2 \log N+\left| A \right|)$ complexity, where $\left| A \right|$ is the number of node-peers between the peers responsible for $k_\ell$, $k_r$ respectively. The theorem follows.

\noindent{\bf Theorem 4}: Range queries of the form $[k_\ell, k_r]$ in the ART structure require an $O(\log_{b}^2 \log N+\left| A \right|)$ complexity, where $\left| A \right|$ is the answer size.

\subsubsection{Query Processing, Data Insertion and Data Deletion, Peer Join and Peer Departure}

In the following we briefly present the basic routines for query processing, data insertion and data deletion, peer join and peer departure.

\begin{algorithm}
\caption{Range\_Search(s,$k_{\ell}$,$k_r$,A)}
\label{alg_search}
\begin{algorithmic}[1]
\STATE Input: $s,k_{\ell},k_r$ (we are at peer $s$ and we are looking for keys in range $\left[k_{\ell},k_r\right]$) 
\STATE Output: $idW$ (the identifier of cluster-peer $W$, which stores $k_{\ell}$ key), $A$ (the answer)
\STATE BEGIN
\STATE We compute $idS$:the identifier of Cluster\_peer $S$, which contains peer $s$;
\STATE We compute $idW$:let $j$ be the identifier of target Cluster\_peer $W$, which stores $k_{\ell}$ key;
\STATE Let $T$ the basic ART structure of cluster-peers;
\STATE W=ART\_Lookup($T,S,idS,W,idW$); \COMMENT{call of the basic routine}
\STATE A=Linear\_Scan of all Cluster\_peers located in and right to $W$ until we find a $key>k_r$;
\STATE END
\end{algorithmic}
\end{algorithm}

The \emph{Range\_Search($s,k_{\ell},k_r$)} routine (Algorithm 1) gets as input the peer $s$ in which the query is initiated and the respective range of keys $\left[k_{\ell},k_r\right]$ and returns as output the $id$ of the cluster\_peer $S$, which contains peer $s$ as well as the cluster\_peer $W$ in which the key $k_{\ell}$ belongs. Then, it calls the basic \emph{ART\_Lookup($T,S,idS,W,idW$)} routine, in order to locate the target peer responsible for key $k_{\ell}$, and then, exploiting the order of keys on each peer performs a right linear\_scan till it finds a $key>k_r$.\\
The \emph{ART\_Lookup($T,S,idS,W,idW$)} routine (Algorithm 2) gets as input the cluster\_peer $S$ (with identifier $idS$) in which the query is initiated and returns as output the $id$ ($idW$) of the cluster\_peer $W$ in which the key $k_{\ell}$ belongs. $T$ denotes the ART-tree structure.
Moreover, Algorithm 2 requires $O(\log_{b}^2 \log N)$ hops, according to first part ($T_1(N)$) of Theorem 3. Obviously, the same complexity holds for insert/delete key operations (see Algorithms 3 and 4), since we have to locate the target peer into which the key must be inserted or deleted. 

For join (depart) peer operations (for details see Algorithm 5), we need $O(\log_{b}^2 \log N) + T_{join}(N)$ ($O(\log_{b}^2 \log N)+T_{depart}(N)$) lookup messages, where $T_{join}(N)$   ($T_{depart}(N)$) is the number of hops required from the respective decentralized structure for peer-join (peer-departure).

\begin{algorithm}
\caption{ART\_Lookup($T,S,idS,W,idW$)}
\label{alg_art-lookup}
\begin{algorithmic}[1]
\STATE Input: We are at cluster-peer $S$ with identifier $idS$
\STATE Output: We are looking for the cluster-peer $W$ with identifier $idW$
\STATE BEGIN
\STATE If ($S$ is responsible for $k_{\ell}$)
\STATE \hspace{2ex} Return $S$;
\STATE Else
\STATE \hspace{2ex} If $W$=1 then $i$=0;
\STATE \hspace{2ex} Else if $W\in\left\{2,3,\ldots, b+1\right]$ then $i$=1;
\STATE Else
\STATE \hspace{2ex} $x$=b+2;
\STATE \hspace{2ex} For ($i=2; i<c_{1}log \log_{b} N; ++i$)
\STATE \hspace{4ex} $x=father(x)+b^{2^{i-2}}$;
\STATE \hspace{4ex} If $j<x$ then break( );
\STATE Follow the RSI[$i$] pointer of cluster\_peer $S$;
\STATE Let $X$ the correspondent cluster\_peer;
\STATE Search for $W$ the 2-level LRT structure starting from $X$;
\STATE Let $Y$ the first cluster-peer of the correspondent collection;
\STATE Let $T'$ the ART structure of the collection above at next level of nesting with root the cluster-peer $Y$;
\STATE $S=Y$;
\STATE ART\_Lookup($T',S,idS,W,idW$); \COMMENT{recursive call of the basic routine}
\STATE Return $W$;
\STATE END
\end{algorithmic}
\end{algorithm}

\begin{algorithm}
\caption{ART\_insert($T,s,k$)}
\label{alg_art-insert}
\begin{algorithmic}[1]
\STATE Input: We are at peer $s$ and we want to insert the key $k$
\STATE Output: The peer $w$ in which $k$ must be inserted
\STATE BEGIN
\STATE We compute $idS$:the identifier of Cluster\_peer $S$, which contains peer $s$;
\STATE We compute $idW$:let $j$ be the identifier of target Cluster\_peer $W$, which stores the $k$ key;
\STATE ART\_Lookup($T,S,idS,W,idW$);
\STATE Let $W$ the target cluster\_peer;
\STATE Search $W$ for peer $w$ containing $k$;
\STATE If $k$ does not exist into $w$, then insert $k$ into it;
\STATE END
\end{algorithmic}
\end{algorithm}

\begin{algorithm}
\caption{ART\_delete($T,s,k$)}
\label{alg_art-delete}
\begin{algorithmic}[1]
\STATE Input: We are at peer $s$ and we want to delete the key $k$
\STATE Output: The peer $w$ in which $k$ must be deleted
\STATE BEGIN
\STATE We compute $idS$:the identifier of Cluster\_peer $S$, which contains peer $s$;
\STATE We compute $idW$:let $j$ be the identifier of target Cluster\_peer $W$, which stores the $k$ key;
\STATE ART\_Lookup($T,S,idS,W,idW$);
\STATE Let $W$ the target cluster\_peer;
\STATE Search $W$ for peer $w$ containing $k$;
\STATE If $k$ exists into $w$, then delete it;
\STATE END
\end{algorithmic}
\end{algorithm}

In the peer join algorithm we assumed that the new peer is accompanied
by a key, and this key designates the exact position in which the new peer
must be inserted. If an empty peer $u$ makes a join request at a
particular peer $v$ (which we call {\em entrance peer}) then there is no
need to get to a different cluster peer than the one in which $u$ belongs.
Similarly, the algorithm for the departure of a peer $u$ assumes that the
request for departure of peer $u$ can be made from any peer in the
ART-structure. This may not be desirable, and in many applications it is
assumed that the choice for departure of peer $u$ can be made only from
this peer. Of course, in this way the algorithm for peer departure is
simplified since there is no need to traverse the ART structure but
only the cluster peer in which $u$ belongs.
In order to bound the size of each cluster\_peer we assume that the
probability of picking an entrance peer is equal among all existing peers,
and that the probability of a peer departing is equal among all existing
peers in the ART. Since the size of the cluster\_peer is bounded by $polylogN$
expected w.h.p., the following theorem is established:

{\bf Theorem 5}: The peer join/departure can be carried out in $O(loglogN)$ hops or lookup messages.

\begin{algorithm}
\caption{ART\_join/leave\_peer($T,s,w$)}
\label{alg_art-join-leave-peer}
\begin{algorithmic}[1]
\STATE Input: We are at peer $s$ and we want to insert/delete the new peer $w$
\STATE Output: The cluster\_peer $W$ in which the peer $w$ must be inserted/deleted
\STATE BEGIN
\STATE We compute $idS$:the identifier of Cluster\_peer $S$, which contains peer $s$;
\STATE We compute $idW$:let $j$ be the identifier of target Cluster\_peer $W$, which contains peer $w$;
\STATE ART\_Lookup($T,S,idS,W,idW$); \COMMENT{call of the basic routine}
\STATE Let $W$ the target cluster\_peer;
\STATE Insert/delete $w$ into/from $W$;
\STATE END
\end{algorithmic}
\end{algorithm}

\subsubsection{Node Failure, Fault Tolerance, Network Restructuring and Load Balancing}

Since we have modeled the join/leave of peers inside a cluster\_peer as the combinatorial game of bins and balls presented in \cite{KMSTTZ03}, each cluster\_peer of an ART structure (according to lemma 4) never exceeds a polylogarithmic number of peers and never becomes empty in expected case with high probability. The latter means that the skeleton ART structure of cluster\_peers remains unchanged in the expected case with high probability as well as in each cluster\_peer the algorithms for peer failure, network restructuring and load balancing are according to the polylogarithmic-sized decentralized architecture we use.

\subsubsection{Multi-attribute Queries}

As in \cite{JOTVZ06}, we divide the whole range of attributes into several sections: each section is used to index an attribute (if it appears frequently in queries) or a group of attributes (if these attributes rarely appear in queries). Since ART can only support queries over one-dimensional data, if we index a group of attributes, we have to convert their values into one-dimensional values (by choosing Hilbert space filling curve or other similar methods). For example, if we have a system with 12 attributes: $a_1,a_2,\cdots,a_{12}$ in which only 4 attributes from $a_1$ to $a_4$ are frequently queried (i.e. 90\% of all queries), we can build 4 separate indexes for them. The remaining attributes can be divided equally into two groups to index, four attributes in each group. This way, the number of replications can be significantly reduced from 12 down to 6.

\section{Evaluation}

For evaluation purposes we used the Distributed Java D-P2P-Sim simulator presented in \cite{SPSTM09}.
The D-P2P-Sim simulator is extremely efficient delivering $>100,000$ cluster peers in a single computer system, using 32-bit JVM 1.6 and 1.5 GB RAM and full D-P2P-Sim GUI support. When 64-bit JVM 1.6 and ~5 RAM is utilized the D-P2P-Sim simulator delivers $>500,000$ cluster peers and full D-P2P-Sim GUI support in a single computer system. When D-P2P-Sim simulator acts in a distributed environment with multiple computer systems with network connection delivers multiple times the former population of cluster peers with only ~10\% overhead.\\
Our experimental performance studies include a detailed performance comparison with BATON*, one of the state-of-the-art decentralized architectures. In particular, we implemented each cluster\_peer as a BATON* \cite{JOTVZ06}, the best known decentralized tree-architecture. We tested the network with different numbers of peers ranging up to 500,000. A number of data equal to the network size multiplied by 2000, which are numbers from the universe [1..1,000,000,000] are inserted to the network in batches. The synthetic
data (numbers) from this universe were produced by the following
distributions: beta, uniform and power-law. For each test, 1,000 exact
match queries and 1,000 range queries are executed, and the average costs
of operations are taken. Searched ranges are created randomly by getting
the whole range of values divided by the total number of peers multiplies
$\alpha$, where $\alpha \in [1..10]$. Note that in all experiments the default value of parameter $b$ is 4.
The source code of the whole evaluation process is publicly available \footnote{\tt
http://code.google.com/p/d-p2p-sim/}.

\subsection{Single- and Multi-attribute Query Performance}

\begin{figure}[h]
\begin{minipage}[b] {0.5\textwidth}
\centering
\includegraphics[width=60mm, height=37mm]{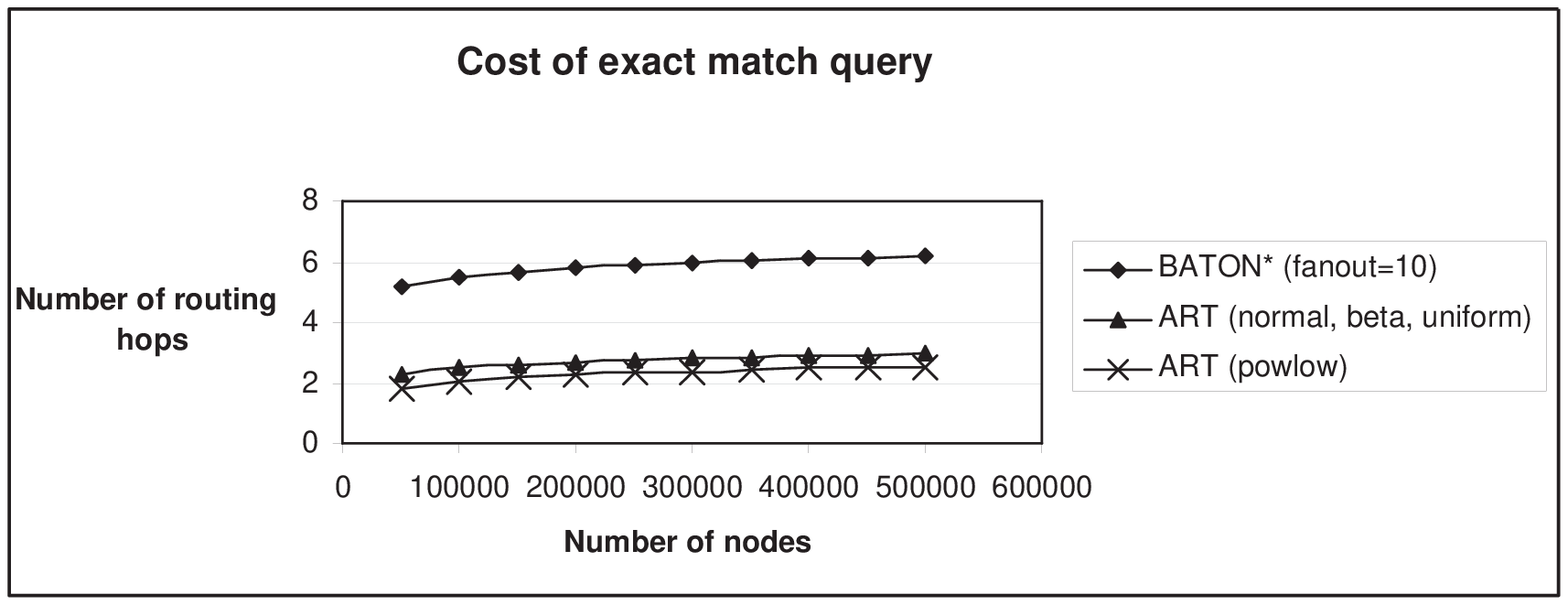}
\end{minipage}
\hfill
\begin{minipage}[b] {0.5\textwidth}
\centering
\includegraphics[width=60mm, height=37mm]{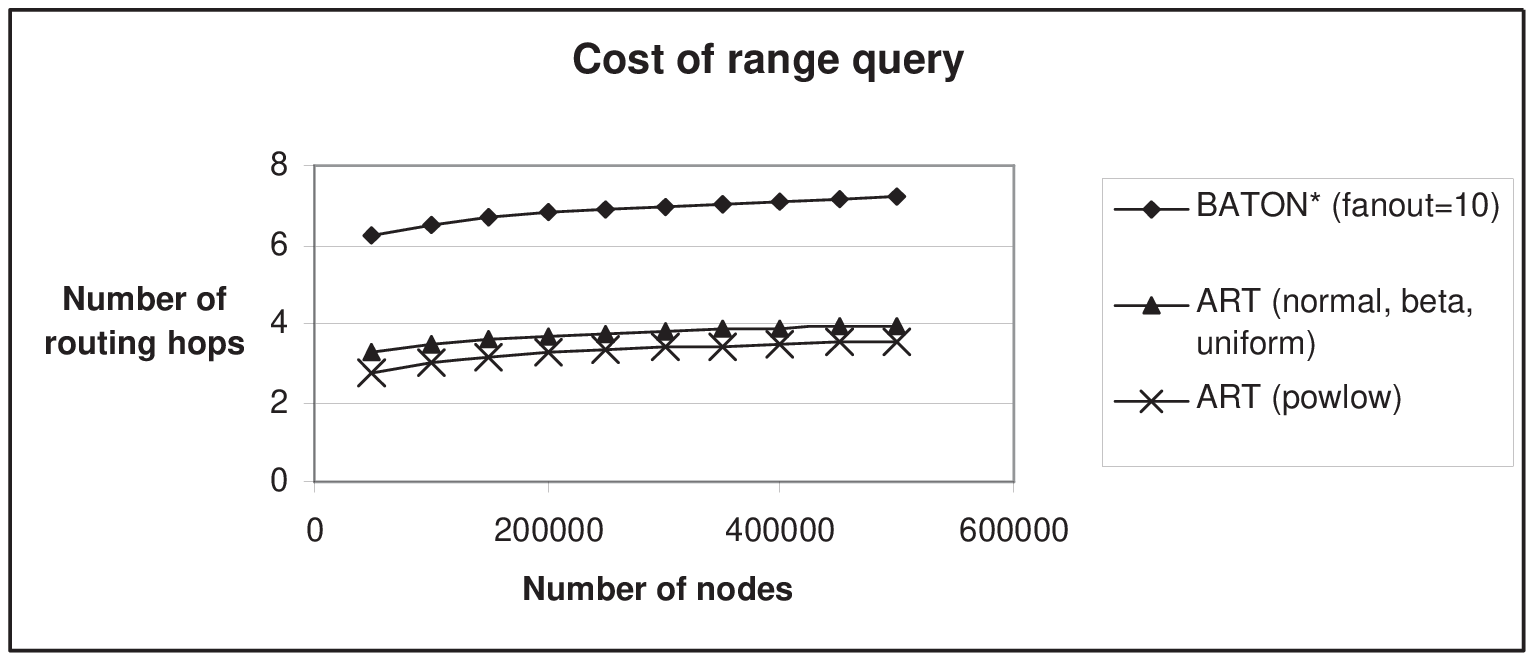}
\end{minipage}
\caption{Cost of exact match query
(left) and cost of range query (right).}
    \label{fig:exact-range(1d)}
\end{figure}

\begin{figure}[h]
\begin{minipage}[b] {0.5\textwidth}
\centering
\includegraphics[width=60mm, height=37mm]{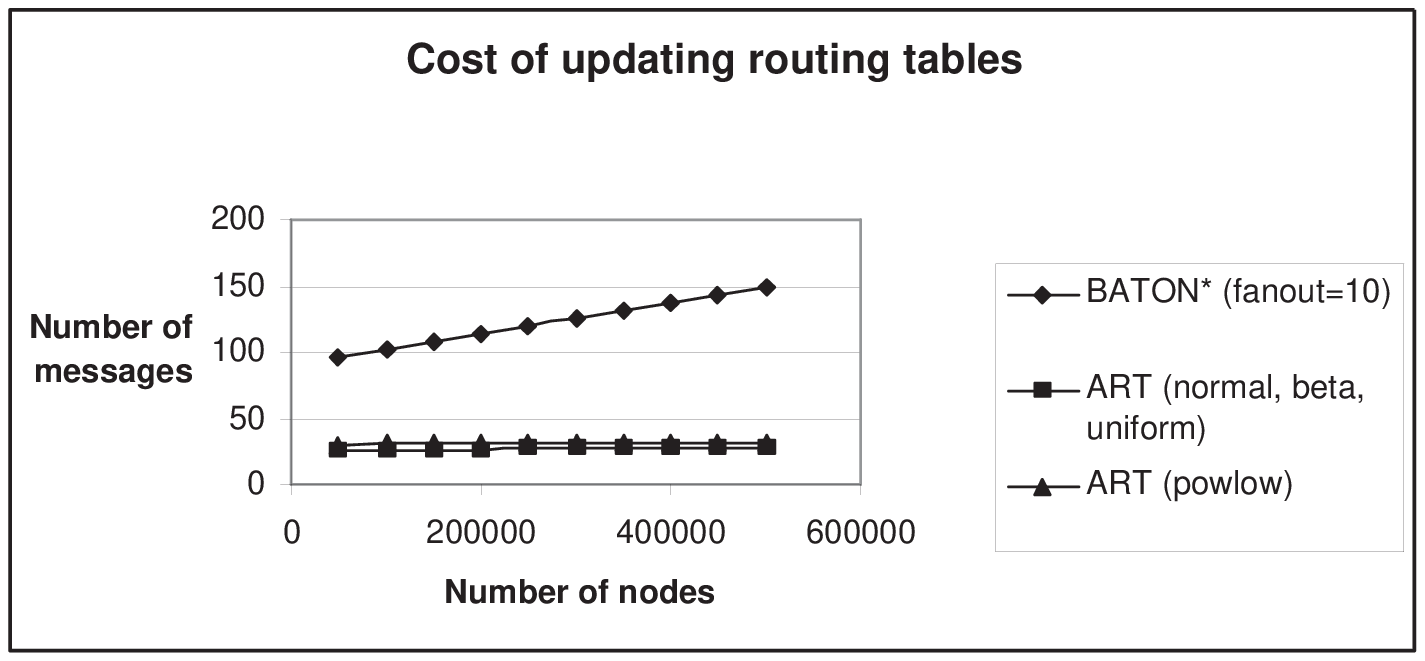}
\end{minipage}
\hfill
\begin{minipage}[b] {0.5\textwidth}
\centering
\includegraphics[width=60mm, height=37mm]{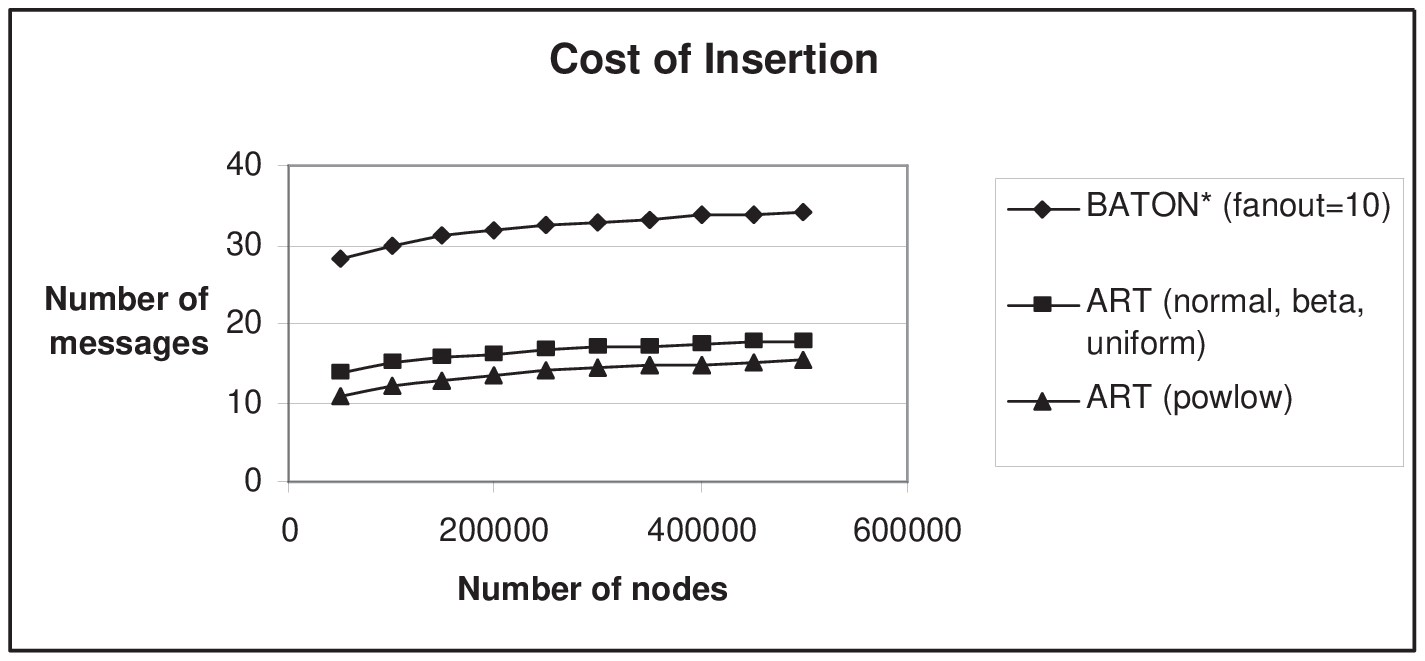}
\end{minipage}
\caption{Cost of updating routing tables
(left) and cost of insertion (right).}
    \label{fig:updaterout-insertion}
\end{figure}

\begin{figure}[h]
\begin{minipage}[b] {0.5\textwidth}
\centering
\includegraphics[width=60mm, height=37mm]{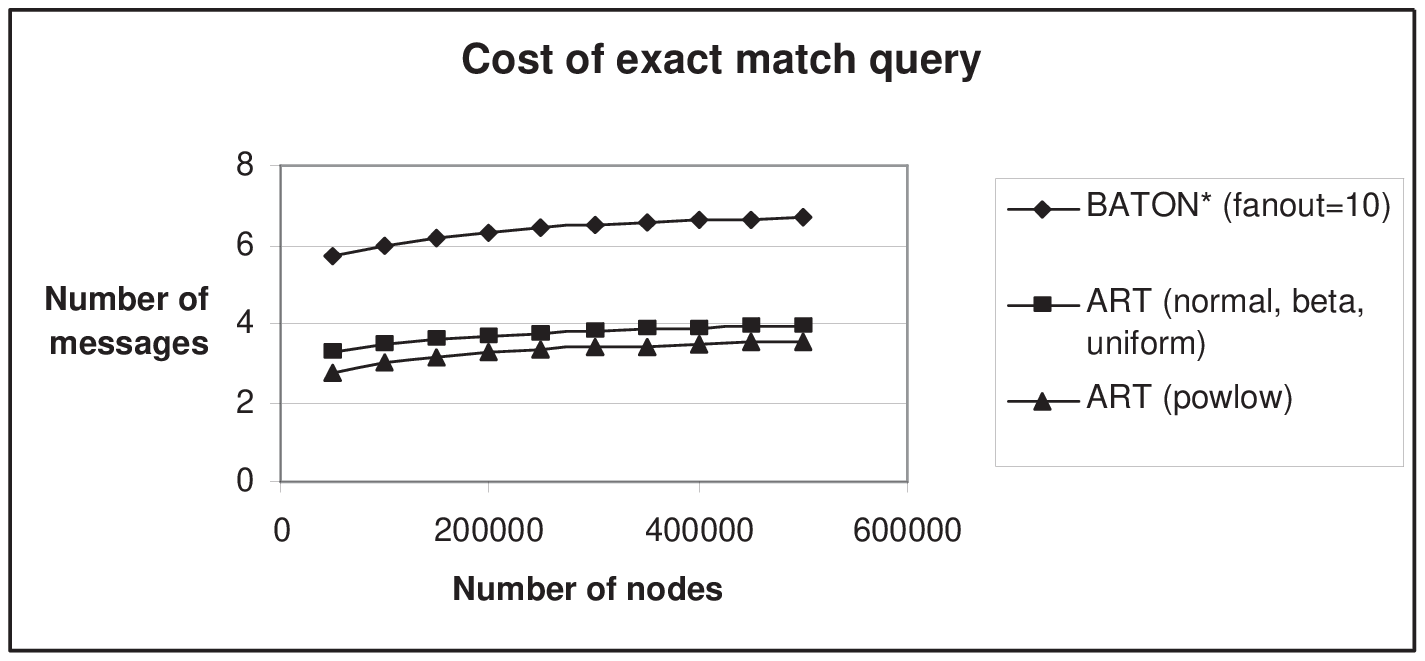}
\end{minipage}
\hfill
\begin{minipage}[b] {0.5\textwidth}
\centering
\includegraphics[width=60mm, height=37mm]{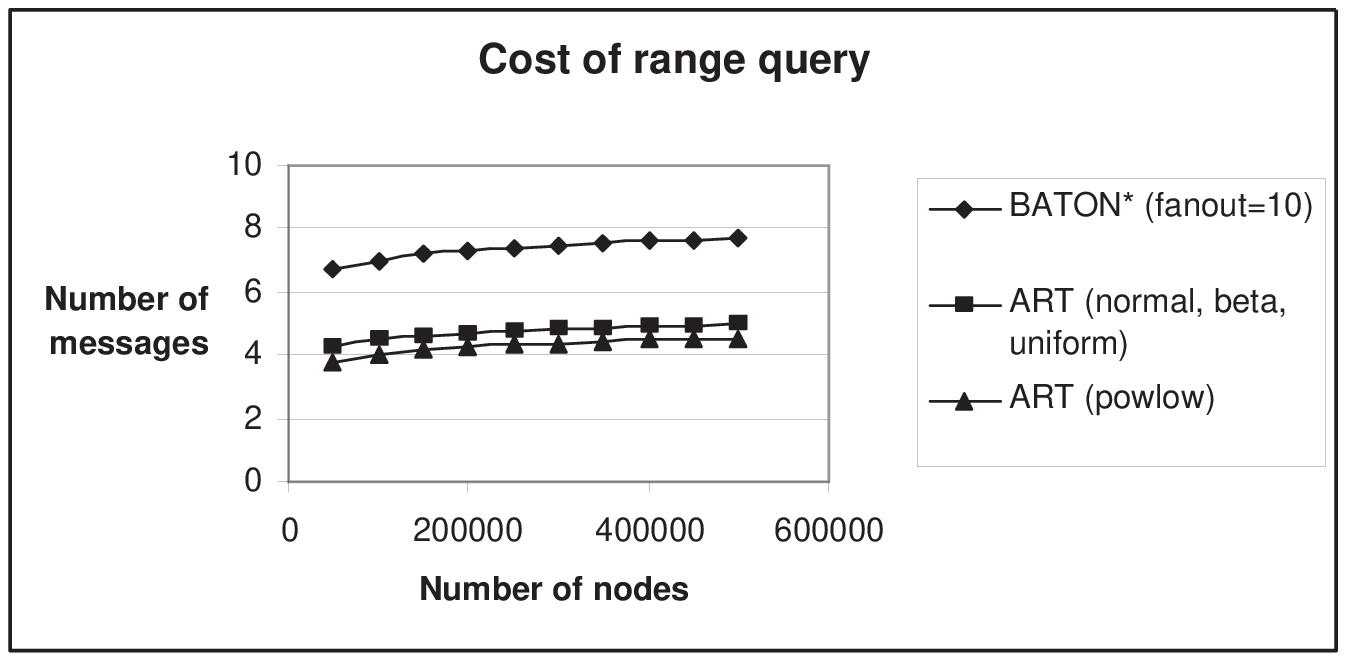}
\end{minipage}
\caption{Cost of multi-attribute exact-match
(left) and range queries (right).}
    \label{fig:exact-range(kd)}
\end{figure}

As proved previously, the whole query performance of ART is $O(\log_{b}^2 \log N')$ where the $N'$ cluster\_peers structure their internal peers according to the BATON* architecture. For normal, beta and uniform distributions each cluster\_peer contains $0.75 \log^2 N$ peers on average and for power-law distributions each cluster\_peer contains $2.5 \log^2 N$ peers on average. Thus, in the former case the average number of cluster\_peers is $N'=\frac{N}{0.75 \log^2 N}$, whereas in the latter case the number of cluster\_peers becomes $N'=\frac{N}{2.5 \log^2 N}$ on average. In all cases, ART outperforms BATON* by a wide margin. As depicted in Figure~\ref{fig:exact-range(1d)} (up), our method is almost 2 times faster and as
a consequence we have a 50\% improvement. The results are analogous with respect to the cost range queries as depicted in Figure \ref{fig:exact-range(1d)} (down).\\
Figure \ref{fig:updaterout-insertion} (up) depicts the cost of updating routing tables. Since each cluster\_peer structures $O(N/{\rm polylog}~N)$ (and not $O(N)$) peers according to BATON* architecture, the results are as expected. We remark that BATON* requires $m \log_m N$ hops, whereas $m \log_m {\rm polylog}~N$ hops are required by ART. In particular and as depicted in Figure \ref{fig:updaterout-insertion} (up), our method updates the routing tables 3 or 4 times faster.
Figure \ref{fig:updaterout-insertion} (down) depicts the insertion cost in multi-attribute case, where we have 6 separate indexes. BATON* requires $6 \log N$ hops and ART requires $6 \log_{b}^2 \log(N/{\rm polylog}~N) + 6 \log ({\rm polylog}~N)$ hops. We observe that the insertion cost of ART is the lowest for any distribution. Again, our method is almost 2 times faster. Finally, the results are analogous for multi-attribute exact-match and range queries respectively (see Figures \ref{fig:exact-range(kd)} (up) and \ref{fig:exact-range(kd)} (down)).

\subsection{Load Balancing}

\begin{figure}[h]
\begin{minipage}[b] {0.5\textwidth}
\centering
\includegraphics[width=60mm, height=37mm]{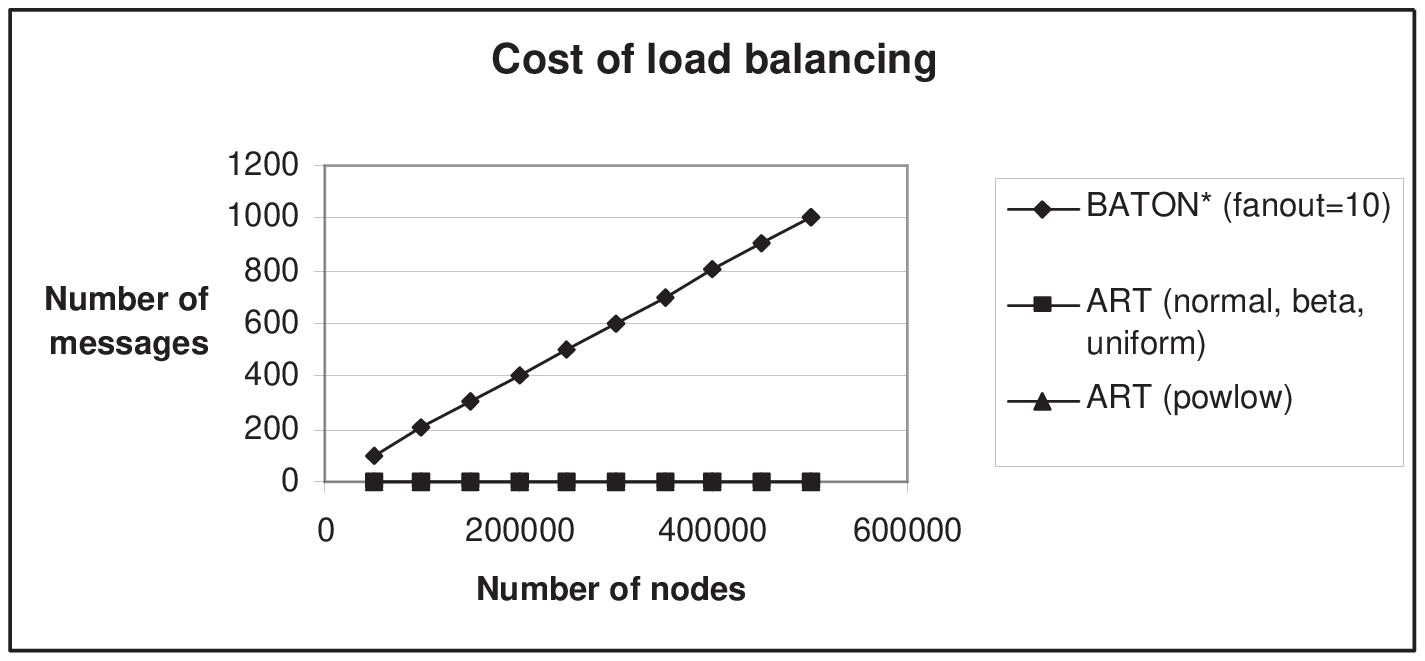}
\end{minipage}
\hfill
\begin{minipage}[b] {0.5\textwidth}
\centering
\includegraphics[width=60mm, height=37mm]{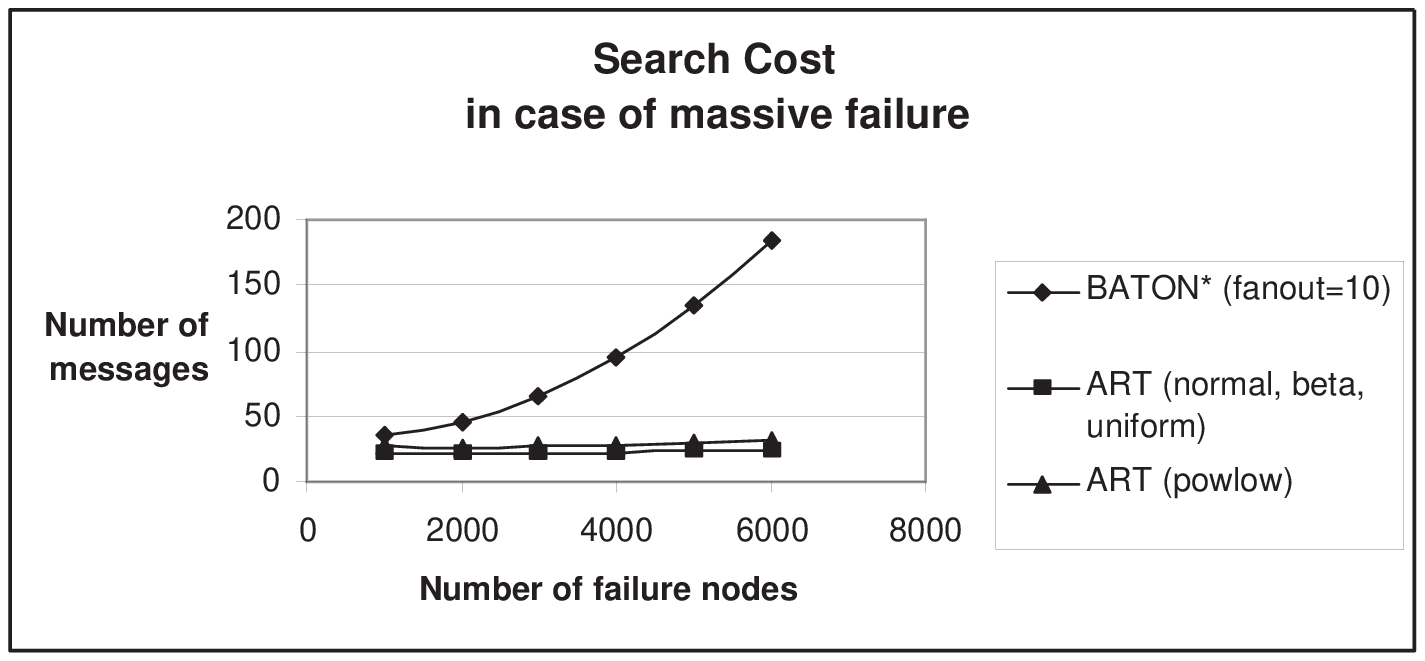}
\end{minipage}
\caption{Cost of load-balancing
(left) and search cost in case of massive failure (right).}
    \label{fig:load-failure(kd)}
\end{figure}

ART not only reduces the search cost but also achieves better load balancing. To verify this claim, we test the network with a variety of distributions and evaluate the cost of load balancing. For simplicity, in our system, we assume that the query distribution follows the data distribution. As a result, the workload of a peer is determined only by the amount of data stored at that peer. In BATON*, when a peer joins the network, it is assigned a default upper and lower load limit by its parent. If the number of stored data at the peer exceeds the upper bound, it is considered as an overloaded peer and vice versa. If a peer is overloaded and cannot find a lightly loaded leaf peer, it is likely that all other peers also have the same work load; thus, it automatically increases the boundaries of storage capability. In ART the overlay of cluster\_peer remains unaffected in the expected case with high probability when peers join or leave the network. Thus, the load-balancing performance is restricted inside a cluster\_peer (which is a new BATON* structure) and as a result ART needs no more than $4$ lookup-messages (instead of $1000$ messages needed from BATON* in case of $500.000$ nodes). For details see Figure \ref{fig:load-failure(kd)} (up).

\subsection{Fault Tolerance}

To evaluate the system's fault tolerance in case of massive failure we initialized the system with 10,000 peers. In the sequel, we let peers randomly fail step by step without recovering. At each step, we check to see if the network is partitioned or not. With massive peer failures, we face a massive destruction of links connected to failed peers. Since the search process has to bypass these peers, the search query has to be forwarded forth and back several times to find a way to the destination and as a result the search cost is expected that will increase substantially. Since the backbone of ART structure remains unaffected w.h.p., meaning that there is always a peer for playing the role of cluster representative, the search cost is restricted inside a cluster\_peer (which is a BATON* structure) and as a result ART needs no more than $32$ lookup-messages (instead of $180$ messages needed from BATON* in case of $6.000$ nodes). Figure \ref{fig:load-failure(kd)} (down) illustrates this effect.

\section{Trade-offs and Heuristics}

If each collection of cluster\_peers is organized
individually as a BATON$^*$ structure (not the whole level of
collections), then we can climb up the ART structure until we reach
the nearest common ancestor of the cluster\_peer we are located in, and
the cluster peer we are searching. Then a downwards traversal is initiated
to reach this cluster\_peer. Since, each collection of $i^{th}$-level is organized according to BATON*, we can decide in $O(\log_m n^{1/2^i})$ hops the child we must follow for further searching. As a result, the total time becomes $O(\log_m n)$ and no improvement has been achieved.

In our solution, if we parameterize the size of the buckets (depicted in Figure 3) from $O(\log^{2c} N)$ to $O(\log^{2f(N)} N)$, where $f(N)$ is a function of the network size, then we can get an interesting trade-off
between the routing data overhead and the number of hops for an operation. In particular, if $Z$ is the number of collections at the current level, then each bucket contains $O(\frac{Z}{\log^{2f(N)}N})$ collections. Thus, the first LRT layer organizes $O(\log^{2f(N)}N)$ bucket representatives and each second LRT layer organizes $O(\frac{Z}{\log^{2f(N)}N})$ collections. In this case, the routing overhead is dominated by the second layer LRTs which becomes $O(\frac{N^{1/4}}{\log^{f(N)}N})$. To achieve an optimal routing data overhead we would like the following: $O(\frac{N^{1/4}}{\log^{f(N)}N})=O(1)\Leftrightarrow f(N)=O(\log N)$. In this case the first LRT layer contains $O(\log^{2f(N)}N)$ or $O(\log^{2\log N}N)$ bucket representative nodes. Therefore, a lookup operation in first layer requires $O(\log \log(\log^{2 \log N}N))$ or $\omega(\log \log N)$ hops. Each of the second layer LRTs contains $O(\frac{Z}{\log^{2f(N)}N})$ collection representative nodes, where $Z$ is the number of collections at current level.
Therefore, the number of hops required by a lookup operation in second layer is $O(\log \log N)$. So, the total time becomes $\omega(\log \log N)$ and the sub-logarithmic complexity is not guaranteed. As a result, if we want an optimal routing overhead we cannot guarantee sub-logarithmic complexity. If we relax the routing overhead to be of polynomial size then we can achieve this.
 
In our solution the routing data overhead ($O(N^{1/4}/ \log^c N)$) is a polynomial function. However, in reality even for an extremely large number of peers N=1.000.000.000, the routing data overhead is 6 for $c=1$,
which is less than the fanout of BATON$^*$ ($m=10$) that we used to run
our experiments. The latter demonstrates the significance of
our result.

\section{Conclusions}

We presented a new efficient decentralized infrastructure for range query processing with probabilistic guarantees, the ART structure. Theoretical analysis showed that the communication cost of query, update and join/leave node operations scale sub-logarithmically expected w.h.p.. Experimental performance comparison with BATON*, the state-of-the-art decentralized structure, showed the improved performance, scalability and efficiency of our new method. Finally, we believe that ART will enable general purpose decentralized trees to support a wider class of queries, and then broaden the horizon of their applicability.

\end{document}